\documentclass[aps,twocolumn,showpacs,superscriptaddress]{revtex4}
\usepackage{graphicx}
\usepackage{psfrag}
\usepackage{color}
\begin{document}
\newcommand{\beq}{\begin{equation}}
\newcommand{\eeq}{\end{equation}}
\newcommand{\ben}{\begin{eqnarray}}
\newcommand{\een}{\end{eqnarray}}
\newcommand{\bea}{\begin{array}}
\newcommand{\eea}{\end{array}}
\newcommand{\om}{(\omega )}
\newcommand{\bef}{\begin{figure}}
\newcommand{\eef}{\end{figure}}
\newcommand{\leg}[1]{\caption{\protect\rm{\protect\footnotesize{#1}}}}
\newcommand{\ew}[1]{\langle{#1}\rangle}
\newcommand{\be}[1]{\mid\!{#1}\!\mid}
\newcommand{\no}{\nonumber}
\newcommand{\etal}{{\em et~al }}
\newcommand{\geff}{g_{\mbox{\it{\scriptsize{eff}}}}}
\newcommand{\da}[1]{{#1}^\dagger}
\newcommand{\cf}{{\it cf.\/}\ }
\newcommand{\ie}{{\it i.e.\/}\ }

\title{Time Scale for Magnetic Reversal at The Topological Non--connectivity Threshold}

\author{G.~L.~Celardo}
\affiliation{Dipartimento di
Matematica e Fisica, Universit\`a Cattolica, via Musei 41, 25121
Brescia, Italy}
\author{J.~Barr\'e}
%\affiliation{Laboratoire de Physique, ENS-Lyon, 46 All\`ee d'Italie, 69364 Lyon
%C\'edex 07, France}
\affiliation{Theoretical Division, Los Alamos National Laboratory, USA}
\author{F.~Borgonovi}
\affiliation{Dipartimento di
Matematica e Fisica, Universit\`a Cattolica, via Musei 41, 25121
Brescia, Italy}
\affiliation{C.N.R.-I.N.F.M.,  Brescia and 
I.N.F.N., Sezione di Pavia, Italy}
\author{S.~Ruffo}
\affiliation{Dipartimento di Energetica ``S.Stecco'' and CSDC,
Universit\`a di Firenze, 
and I.N.F.N., via Santa Marta 3, 50139, Firenze, Italy}

\begin{abstract}

Anisotropic classical Heisenberg models with all-to-all spin coupling
display a Topological Non--connectivity Threshold (TNT) for any number
$N$ of spins. Below this threshold, the energy surface is disconnected in two
components with positive and negative total magnetizations respectively,
so that magnetization cannot reverse its sign and ergodicity is broken,
even at finite $N$. Here, we solve the model in the microcanonical
ensemble, using a recently developed method based on large deviation
techniques, and show that a phase transition is present at an energy
higher than the TNT energy. In the energy range between the TNT energy
and the phase transition, magnetization changes sign
stochastically and its behavior can be fully characterized by an
average magnetization reversal time.  The time scale for magnetic
reversal can be computed analytically, using statistical mechanics.
Numerical simulations confirm this calculation and further show that
the magnetic reversal time diverges with a power law at the TNT threshold,
with a size dependent exponent. This exponent can be computed in the
thermodynamic limit ($N\to\infty$), by the knowledge of  entropy
as a function of magnetization, and turns out to be in reasonable
agreement with finite $N$ numerical simulations.
We finally generalize our results to other models: Heisenberg chains
with distance dependent coupling, small 3D clusters with nearest
neighbor interactions, metastable states. We conjecture that the
power-law divergence of the magnetic reversal time scale might be
a universal signature of the presence of a TNT.

\end{abstract}
\date{\today}
\maketitle

\section{Introduction}

Statistical mechanics deals with systems containing a very large
number ($10^{23}$) of interacting particles. Nowadays, as the
experimental investigation of few-atom systems is becoming
possible, the analysis of small systems raises fundamental
questions~\cite{Gross}, and the problem of a statistical description
of few-body systems with strong nonlinear interaction is a subject of
current research~\cite{chaos}. Unfortunately we are still far from
understanding what are the conditions for a few-body system to reach,
if any, an equilibrium, and how to describe it in the same way as
statistical mechanics provides a powerful description of
large systems.  For instance, even the existence of
temperature at the nanoscale has been recently questioned
in Ref.~\cite{nano}.

It has been recently shown that, for a Heisenberg model with
all-to-all coupling, there exists a specific energy threshold below
which total magnetization cannot change its sign, even when the number
of spins is finite~\cite{jsp}. This ergodicity breaking phenomenon has
been related to the existence of a Topological Non-connectivity
Threshold (TNT) of the energy surface.  
%On the other hand, above the
%energy threshold, as we will show in this paper, the dynamics can be
%characterized by a mean reversal time.  
This type of ergodicity
breaking at finite $N$ is different from the $N\to \infty$ ergodicity
breaking in a standard Ising model below the critical temperature.
  The existence of this
threshold is not restricted to the infinite range coupling case. It is
also present when the interaction among Heisenberg spins decays as
$R^{-\alpha}$, where $R$ is the distance between two spins.  It has
been indeed proved~\cite{brescia} that, for a $d-$dimensional system,
the ratio of the disconnected portion of the energy range with respect
to the total energy range tends to zero in the thermodynamic limit for
$\alpha>d$ (short range interactions) while it remains finite for
$\alpha<d$ (long range interactions).  On the other hand, although the
mean-field (all-to-all) type of spin coupling might appear unphysical,
magnetic systems can be realized, using modern experimental techniques
\cite{cornell}, which are well described by Heisenberg--like
Hamiltonians with an infinite range term.  Moreover, when the range of
the interaction is of the same order of the size of the system,
all-to-all coupling may be a meaningful first order
approximation~\cite{Gross,thebibble}. This could be the case for small
systems used in current nano-technology, which requires to deal with
systems made of a few dozens of particles. Otherwise, all-to-all
coupling is relevant for macroscopic systems with long range
interactions, like gravitational and unscreened Coulomb
systems~\cite{thebibble}.
  
We address in this paper the issue of providing a theoretical
framework to calculate the magnetization  reversal time for the
mean-field anisotropic Heisenberg model in a magnetic field, in
which a finite number $N$ of spins interact with all-to-all couplings.
We have already stressed that below the TNT the total magnetization
cannot reverse its sign, thus  magnetization does not relax to its
equilibrium value. In this paper we will answer the following
questions: $i)$ Above the TNT, does the magnetization reverse its sign? 
If so, on which time scale?  $ii)$ What is the
relevance of the TNT in a system with a standard magnetic phase
transition? Thus, the aim is to explain the main physical effects
associated with the TNT, and how it affects the
phase transition appearing in this model at a higher energy. The
latter is studied in the microcanonical ensemble, applying a recently
developed solution method of mean-field Hamiltonians based on
large-deviation theory~\cite{thierry}. We study in detail, numerically
and analytically, the time scale for magnetization reversal.  At the
TNT, the reversal time diverges as a power law, with a characteristic
exponent proportional to the number of spins $N$. Based on analytical
calculations, we expect this property to be universal. Finally, we
extend some of the results obtained for all-to-all coupling to other
models: chains with distance dependent couplings and small clusters
with nearest neighbor interactions.  We show the existence of the TNT
also in these cases, and we present strong evidence for the power law
divergence of  the reversal time.

\section{The Model}
The Hamiltonian of the model is
\begin{equation}
\label{ham}
H=B\sum_{i=1}^{N} S^z_i+ \frac{J}{2} \sum_{i=1}^N 
\sum_{j\ne i} (S_i^x S_j^x - S_i^y S_j^y), 
\end{equation}
where $\vec S_i=(S_i^x,S_i^y,S_i^z)$ is the spin vector with continuous
components, $N$ is the number of  spins, $B$ is the rescaled
external magnetic field strength and $J$ the all-to-all coupling strength
(the summation is  extended over  all pairs).
Let us also  define 
$$
m_{x,y,z}=\frac{1}{N} \sum_{i=1}^N  S_i^{x,y,z},
$$ 
as the components of the total magnetization of the system. 
Due to the anisotropy of the
coupling, the system has an easy--axis of the magnetization 
along the $y$ direction
(the easy--axis of the magnetization is defined
by the direction of the magnetization in the minimal energy configuration
of the system).
The equations of motion are derived in a standard way from Hamiltonian 
(\ref{ham}), and we obtain:

\begin{equation}
\label{eom2}
\left\{ \begin{array}{ccc}
\dot{S}_i^x  &=& -BS_i^y -JS_i^z \sum_{\langle j\rangle} S_j^y\\
\\
\dot{S}_i^y &=& BS_i^x -JS_i^z \sum_{\langle j\rangle} S_j^x\\
\\
\dot{S}_i^z &=& J \sum_{\langle j\rangle} (S_i^yS_j^x+S_i^xS_j^y).
\end{array}\right.
\end{equation}
The total energy $E=H$ and the spin moduli $|\vec {S}_i|^2=1$ are
constants of the motion. 
Dynamics, already studied in a similar model~\cite{jsp,num}, 
is characterized by chaotic motion (positive
maximal Lyapunov exponent) for not too small energy values and spin
coupling constants. For $J=0$ the model is exactly integrable,
while for generic $J$ and $B$ there is a mixed phase space 
with  prevalently chaotic motion for $|E| \lesssim  JN$.

\section{The two thresholds}
\label{sec:thre}

We will now show the existence of two distinct thresholds in this
model: first we  derive analytically the Topological Non--connectivity
Threshold (TNT), then we will present the microcanonical analysis and the
analytical evaluation of the statistical threshold, at which a second
order phase transition occurs in the $N \rightarrow \infty$ limit.

\subsection{The Topological Non--connectivity Threshold}

The phase space of the system is topologically disconnected below
a given energy density $\epsilon_{dis}$, which can be obtained 
as in Ref.~\cite{jsp,phd}. From symmetry
considerations, both positive and negative regions of $m_y$ exist on
the same energy surface.  Indeed the Hamiltonian is invariant under a
rotation of $\pi$ around the $z$ axis for which $S_i^y \rightarrow
-S_i^y$ and $S_i^x \rightarrow -S_i^x$.  
Switching dynamically from a negative $m_y$ value to a
positive one requires, for continuity, to pass through $m_y=0$. Hence,
for all energy values above 
$$\epsilon_{dis}=\mbox{Min}[H/N \ | \ m_y=0]$$
magnetization reversal is possible, while below this value
magnetization cannot change sign. 

Hamiltonian (\ref{ham}) can be written as follows:
\begin{equation}
H= BNm_z+\frac{J}{2}N^2\left( m_x^2-m_y^2 \right)+\frac{J}{2}\sum_i (S_i^y)^2-(S_i^x)^2.
\label{Hamiltonien2}
\end{equation}
The Topological Non-connectivity Threshold (TNT) is defined as the minimum of
the Hamiltonian under the $N+1$ constraints:
\begin{eqnarray}
&&a) \,(S_i^x)^2+(S_i^x)^2+(S_i^x)^2=1 \label{Constra}\\
&&b) \, m_y=0. \label{Constrb}
\end{eqnarray}
Instead of solving the constrained problem,
we simplify it by calculating the absolute minimum of
$$F=BNm_z-\frac{J}{2} \sum[ (S_i^x)^2-(S_i^y)^2].$$
If the minimal solution satisfies both $m_x=0$ and $m_y=0$, 
the problem is equivalent to the original one.
Conditions (\ref{Constra}) are taken into account setting:
$$
S_i^z=\cos\theta_i,\ S_i^x=\sin\theta_i \cos\phi_i,
\ S_i^y=\sin\theta_i \sin\phi_i~.
$$
Taking the derivatives of $F$ we obtain:
\begin{eqnarray}
\frac{\partial{F}}{\partial\phi_i}&=&J\sin^2\theta_i\cos\phi_i\sin\phi_i=0
\label{sup}\\
\frac{\partial{F}}{\partial\theta_i}&=&\sin\theta_i
[B+J\cos\theta_i\cos^2\phi_i]=0~.
\label{deri}
\end{eqnarray}
If $B > J$, Eq.~(\ref{deri}) has the  solution, $\sin\theta_i=0$, that also
satisfies Eq.~(\ref{sup}). It corresponds to all spins lying along 
the $z$-axis and
\begin{equation}
\epsilon_{dis}=-B~.
\label{NE1}
\end{equation}
If $B < J $, then from Eq.~(\ref{deri}) we have two possible solutions
for each~$i$:
\begin{itemize}
\item [1)] $\theta_i=\pi$;
\item [2)] $\sin\phi_i=0$ and $\cos\theta_i=-B/J$
\end{itemize}
Let us define $0\leq n_z\leq N$  as the number of spins satisfying condition 1)
above. Then
$$
F(n_z)=\frac{n_z}{2J}\left(B-J\right)^2
-N\left( \frac{B^2}{2J}+\frac{J}{2}\right),
$$
so that the minimum is reached for $n_z=0$ or $\cos\theta_i=-B/J$ and
$\sin\phi_i=0$ for all $i$. This in turn implies $m_y=0$
and, for $N$ even, $m_x=0$ (choosing for instance $\phi_i=\pi/2$
for $i=1,N/2$ and $\phi_i=-\pi/2$ for $i=N/2+1,N$).
Then we have (for $N$ even):
\begin{equation}
\epsilon_{dis}=-\left(\frac{B^2}{2J}+\frac{J}{2}\right)~.
\label{NE2}
\end{equation}

Summarizing, we get~\cite{odd},
\begin{equation}
\epsilon_{dis} = \left\{ 
\begin{array}{lll}
\displaystyle
&-B ~{\rm for}~J \le B~\\
& \\
\displaystyle
&-(\frac{B^2}{2J}+\frac{J}{2})~{\rm for}~J>B.
\end{array}
\right.
\label{cri}
\end{equation}

The existence of $\epsilon_{dis}$ does not represent a sufficient
condition in order to demagnetize a sample for $\epsilon >
\epsilon_{dis}$.  As it will be shown in Sec.~\ref{sec:quasint}, regular
structures indeed appear in some cases, preventing most
trajectories to cross the $m_y=0$ plane.

\subsection{The Statistical Threshold: phase transition}
\label{sec:stat}

We now determine the statistical phase-transition energy of the
model in the microcanonical ensemble.
To keep the calculations easy, we will first neglect
the term $J/2\sum_i (S_i^y)^2-(S_i^x)^2$ in (\ref{Hamiltonien2}).
We will show later how to take into account this term.
In order to facilitate the calculations, we will also set

\begin{eqnarray}
\epsilon & \rightarrow &\epsilon/B  \nonumber \\
I& \rightarrow &\frac{JN}{B}~.
\label{trasf}
\end{eqnarray}

Thus we can consider the following  Mean-Field Hamiltonian:

\begin{equation}
H_{MF}=N\left[ m_z+\frac{I}{2}\left( m_x^2-m_y^2 \right)\right],
\label{Hamiltonien3}
\end{equation}

Note that this mean field limit is formally identical to phenomenological
single spin Hamiltonians used to model micromagnetic systems
\cite{chud}.

Using this simplified Hamiltonian, we can calculate the entropy,
counting the number of microscopic configurations associated with
given values of $m_x$, $m_y$ and $m_z$, independently of the energy of
the system. This can be done using Cram\'er theorem, a basic tool
of Large Deviation Theory~\cite{Dembo}.  Each single spin is
characterized by two angles $\theta$ and $\phi$, such that $S_z=\cos
\theta$, $S_x=\sin \theta \cos \phi$, $S_y=\sin \theta \sin \phi$. We
calculate the function

\begin{eqnarray}
\Psi(\lambda_x,\lambda_y,\lambda) &=& \frac{1}{4\pi}\int_0^{\pi} \sin
  \theta \; d\theta \int_0^{2\pi} d\phi~e^{\lambda\cos \theta}
  \nonumber \\
 && e^{\lambda_x \sin \theta \cos \phi +\lambda_y \sin \theta \sin \phi}~.
\label{eqpourPsi}
\end{eqnarray}

We then get the entropy $s(m_x,m_y,m_z)$ through a Legendre-Fenchel
transform of $\ln \Psi$:

\begin{eqnarray}
s(m_x,m_y,m_z) &=&-\sup_{\lambda_x,\lambda_y,\lambda}\left[\lambda_x m_x 
+\lambda_y m_y +\lambda m_z \right. \nonumber \\  
&&  \left. -\ln \Psi(\lambda_x,\lambda_y,\lambda) \right]~.
\label{eqpours1}
\end{eqnarray} 

This calculation gives us an approximate expression for the
probability $P(m_x,m_y,\epsilon)$, which describes the system for each
energy:
\begin{equation}
P(m_x,m_y,\epsilon)\propto \exp{(Ns(m_x,m_y,m_z=\epsilon-\frac{I}{2}(m_x^2-m_y^2)))}~.
\end{equation}

Integrating over $m_x$, one gets the marginal probability distribution
$P(m_y, \epsilon)$. We define the paramagnetic (resp. ferromagnetic)
phase by a probability distribution $P(m_y, \epsilon)$ which is single
peaked around $m_y=0$ (resp. double peaked). To locate the statistical
phase transition energy $\epsilon_{stat}$, we assume that the transition is
second order; it is then sufficient to study the entropy around
$m_y=0$.  We will also set $m_x=0$, since it is easy to see that a
non-zero $m_x$ would only decrease the entropy for negative energy
states; thus these states with non zero $m_x$ have little influence.
Physically, the picture is the following: the negative energy has to
be absorbed by either a non-zero $m_z$, or a non-zero $m_y$ or both.
For small negative energies, it is entropically favorable to decrease
a bit $m_z$, since it has a linear effect on the energy. For negative
enough energies however, it costs much entropy to decrease $m_z$
further, so that a non-zero $m_y$ is  favored, this is the phase
transition. As a small $m_y$ results in a small $\lambda_y$, we
develop $\Psi$ and $\ln \Psi$ up to second order in $\lambda_y$:
\begin{eqnarray}
\Psi(\lambda_y,\lambda) & \simeq & \frac{\sinh \lambda}{\lambda}+
\frac{\lambda_y^2}{2}\frac{\lambda \cosh \lambda -\sinh \lambda}{\lambda^3}
\\
\ln \Psi(\lambda_y,\lambda) & \simeq & 
\ln \left(\frac{\sinh \lambda}{\lambda}\right) +\frac{\lambda_y^2}{2}
\frac{\lambda \cosh \lambda-\sinh \lambda}{\lambda^2 \sinh \lambda}~.
\end{eqnarray}
The maximization over $\lambda$ and $\lambda_y$ yields the equations:
\begin{eqnarray}
\label{eq:lambda}
m_z &=& \lambda \phi(\lambda) + \frac{\lambda_y^2}{2} \phi^\prime(\lambda) \\
m_y &=& \lambda_y \phi(\lambda)~, 
\end{eqnarray}
where
\begin{equation}
\label{ento2}
\phi(\lambda) = \frac{\lambda\cosh\lambda-\sinh\lambda}
{\lambda^2\sinh\lambda}~.
\end{equation}
From Eq.~(\ref{eq:lambda}), we  write $\lambda=\lambda_0
+a_2\lambda_y^2$, where $\lambda_0$ is defined implicitly by
\begin{equation}
\label{ento3}
m_z= \lambda_0 \phi(\lambda_0)~,
\end{equation}
and $a_2$ is a coefficient. We then compute the entropy up to second
order in $m_y$:
\begin{equation}
\label{ento1a}
s(m_y, \epsilon) = -\frac{m^2_y}{2\phi(\lambda)} -\lambda_0 \left( m_z
\right) + \ln \left( \frac{\sinh \lambda_0}{\lambda_0} \right)~;
\end{equation}
note that the terms with $a_2$ canceled. Using energy
conservation $m_z=\epsilon +I m_y^2/2$, we obtain the entropy
$s(m_y;\epsilon)$ as a function of $m_y$ alone, $\epsilon$ being now a
parameter. The equation for $\lambda_0$ is:
\begin{equation}
\epsilon +\frac{I}{2} m^2_y = \lambda_0 \phi(\lambda_0)~. 
\end{equation}
We then write $\lambda_0=\mu+m^2_y \mu^2$, with
$\epsilon=\mu\phi(\mu)$, and substitute into
Eq.~(\ref{ento1a}), to get $s(m_y;\epsilon)$ up to second order in $m_y$:
\begin{equation}
\label{ento1}
s(m_y, \epsilon) =  -\mu \epsilon + \ln \left( \frac{\sinh \mu}{\mu} \right)
-m^2_y \left( \frac{1}{2\epsilon}+\frac{I}{2}\right)~.
\end{equation}

The vanishing of the second derivative in $m_y=0$ yields the critical
energy: $\epsilon_{stat}=-1/I$, which can be expressed in the
old variables, see Eq.~(\ref{trasf}):

\begin{equation}
\epsilon_{stat}=-\frac{B^2}{JN}.
\label{estat}
\end{equation}

At this threshold entropy has a maximum in $m_y=0$, with vanishing
second derivative. In the thermodynamic limit the second derivative of
the entropy as a function of $\epsilon$ becomes discontinuous in
$\epsilon_{stat}$, indicating that a true second order phase
transition occurs at $\epsilon_{stat}$, for $N\to\infty$. This
analytically calculated value of $\epsilon_{stat}$ is in reasonable
agreement with numerical results obtained using the full Hamiltonian
(\ref{ham}).

\begin{figure}
\begin{center}
\includegraphics[scale=0.5]{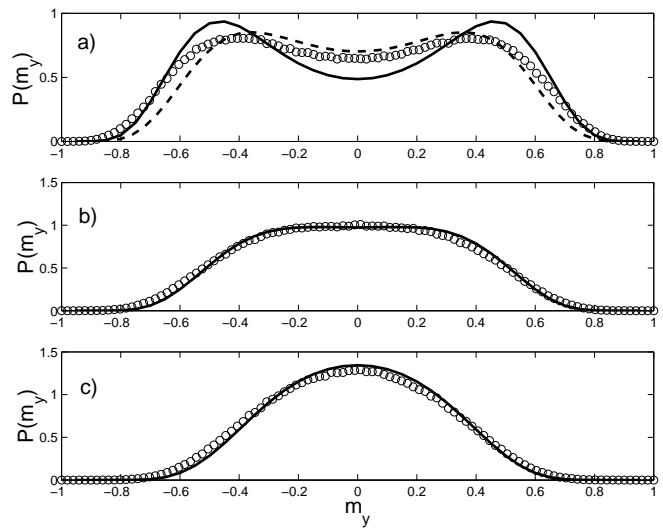}
\caption{Comparison of the distributions $P(m_y)$ obtained
  analytically (solid lines) and numerically (open circles). All plots are
  for $N=6$ spins, a coupling $J=1/3$ and a field $B=1$. From top to
  bottom, the energy per spin is $\epsilon=-0.7$ (ferromagnetic
  phase), $\epsilon=-0.5$ (close to the phase transition), and
  $\epsilon=-0.3$ (paramagnetic phase). In panel $a)$ the dashed line
  shows the improved analytical calculations in which the non-mean 
  field terms are taken into account.
} 
\label{fig:distrib}
\end{center}
\end{figure}

The corresponding probability distribution  $P(m_y,\epsilon)$ obtained
from the mean field Hamiltonian (\ref{Hamiltonien3}), should be compared
with that obtained (numerically) from the full Hamiltonian (\ref{ham}),
for instance by  sampling of the phase space. Results are shown in
Fig.~\ref{fig:distrib}. As one can see the agreement is
quantitatively good in the paramagnetic phase, but only qualitatively
correct in the ferromagnetic phase; here  the double peaked shape is
correct, but the details are significantly off.\\

The inaccuracy of the calculation may come from both the small value of
$N$, and from the term $J/2\sum_i (S_i^y)^2-(S_i^x)^2$, which has been
neglected till now. It can be included in the statistical analysis as
follows. The Hamiltonian depends now on another global quantity,
$\Delta=\langle (S_i^x)^2-(S_i^y)^2\rangle$.  It is possible to
include it in the large deviation calculation; Eq.~(\ref{eqpourPsi}) is
modified into:
\begin{eqnarray}
&&\tilde{\Psi}(\lambda_x,\lambda_y,\lambda,\mu) = \frac{1}{4\pi}\int_0^{\pi} \sin
  \theta \; d\theta \int_0^{2\pi} d\phi \ e^{\lambda\cos \theta}\cdot
  \nonumber \\
  &&e^{\lambda_x \sin \theta \cos \phi +\lambda_y \sin \theta \sin
  \phi} \  e^{\mu\sin^2 \theta (\cos^2 \phi-\sin^2 \phi)}.
\label{eqpourPsi2}
\end{eqnarray}

One proceeds by writing a probability distribution
$P(m_x,m_y,m_z,\Delta) \propto \exp(Ns(m_x,m_y,m_z,\Delta))$, taking
into account the energy conservation $m_z= \epsilon -I(m_x^2-m_y^2)/2
+J\Delta/2$ (for $B=1$), and integrating over $m_x$ and $\Delta$ to
obtain $P(m_y,\epsilon)$. This last step has to be carried out
numerically, and no simple expression as~(\ref{estat}) is available
any more. A comparison with a numerical investigation of the phase
space shows that the additional term has a significant contribution;
the $P(m_y)$ we obtained in the ferromagnetic phase improves on
the mean field calculation, see Fig.~\ref{fig:distrib}.
We conclude that the remaining discrepancies come from the small value
of $N$ ($N=6$ on Fig.~\ref{fig:distrib}).\\

Let us finally remark that in this statistical framework, the TNT
energy, $\epsilon_{dis}$, can be recovered as the
energy such that $s(0,\epsilon_{dis})= -\infty$.  From
Eqs.~(\ref{ento3},\ref{ento1}) with $m_y=0$, it is easy to get
$\epsilon_{dis} = -1$; this implies, using Eq.~(\ref{trasf}),
$\epsilon_{dis} = -B$, which is the same as in Eq.~(\ref{cri}) in the
limit $N\to\infty$ for $J<B$.

\section{Time Scale for Magnetic Reversal}
\label{sec:time}
In the following, we will study the dynamics of the full
Hamiltonian (\ref{ham}), which, at variance with (\ref{Hamiltonien3}), 
is non--integrable and can display chaotic motion.
Let us first notice that in the large $N$ limit 
the minimal energy can be easily estimated 
(see Appendix~\ref{appa}) as

\begin{equation}
\epsilon_{min}=\left\{
\begin{array}{lll}
\displaystyle&-\frac{B^2}{2JN}-\frac{JN}{2} \ {\rm~~~for} \qquad J\geq \frac{B}{N}\\
&\\
\displaystyle&-B \quad{\rm~~~~~~~~for} \qquad J < \frac{B}{N}.
\end{array}
\right.
\label{miesb}
\end{equation}

From Eqs.~(\ref{miesb},\ref{estat},\ref{cri}) we have that if $J>B/N$ then
$\epsilon_{stat} > \epsilon_{dis}> \epsilon_{min}$.
In what follows we will restrict our consideration to the 
region of parameters for which these three thresholds are different.

The two thresholds, $\epsilon_{dis}$ and $\epsilon_{stat}$, define
three energy regions which show different dynamical and statistical
properties:

\begin{figure}
\begin{center}
\includegraphics[scale=0.35]{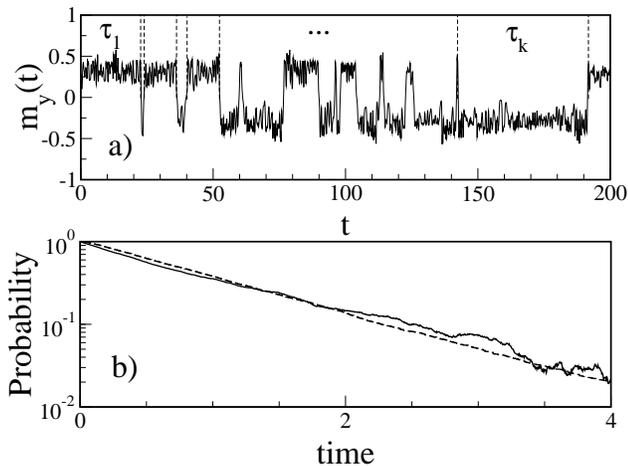}
\caption{
In this figure, all data refer to the $N=6$, $B=1$, $J=3$ case.
a) Magnetization $m_y$ {\it vs} time. Reversal times (times between
neighbor zeros of $m_y$ have been indicated as $\tau_1,\ldots,\tau_k$. 
b) Solid line : probability distribution of reversal times versus
the normalized  time $\tau/\langle \tau\rangle$, for $1$ trajectory
and $10^4$ different crossings.
Dashed  line : probability distribution of relaxation times  
$P=\frac{P_+(t)-1/2}{P_+(0)-1/2}$ {\it vs} $2t/\langle \tau\rangle$.
As initial condition we choose  $P_+(0)=1$ and $10^4$ different 
initial conditions. 
}
\label{ptcl}
\end{center}
\end{figure}

\noindent
1) For $ \epsilon < \epsilon_{dis}$, the probability distribution of
$m_y$ $P(m_y)$ has two separate peaks, with $P(m_y=0)=0$, so that
$m_y$ cannot change sign in time.

\noindent
2) For $\epsilon_{stat}\le\epsilon\le 0$, $m_y$ quickly changes sign
in time and $P(m_y)$ is peaked at $m_y=0$.

\noindent
3) For $\epsilon_{dis}< \epsilon < \epsilon_{stat}$, the probability
distribution is doubly peaked around the most probable values of the
magnetization. These two peaks are not separated and 
$P_0 \equiv P(m_y=0)\ne 0$.
What actually happens dynamically depends on
the relative strength of the coupling $J$ with respect to $B$.
More specifically we can characterize two different behaviors,
chaotic and quasi-integrable.

\subsection{Chaotic Regime}

\subsubsection{Time scale for magnetic reversal and relaxation}
\label{subtime}

For $J$ big enough (fully chaotic regime) the behavior of $m_y(t)$
resembles a random telegraph noise~\cite{tel}, Fig.~\ref{ptcl}a):
magnetization switches stochastically between its two most probable values,
reversing its sign at random times.
If we sample the magnetization reversal times, $\tau_k$, defined as the time
interval between two crossings of $m_y=0$, we find that they follow
a Poissonian distribution with average $\langle \tau \rangle$.
Such  distribution of the reversal times is a consequence of strong chaos:
the system looses its memory due to sensitivity to initial conditions
and the reversal probability per unit time,  $\lambda=1/\langle \tau \rangle $,
becomes time independent.

Since the magnetization reverses its sign randomly, 
any initial macroscopic sample
with $m_y \ne 0$, will relax to an equilibrium distribution with a vanishing
average magnetization.
In order to characterize quantitatively the relaxation process, we introduce 
the probability to have a positive magnetization, $P_+(t) $, at time $t$.
This is measured by considering an ensemble of $n$ initial conditions and 
counting, for each time $t$, the number of trajectories $n_+(t)$ 
for which $m_y>0$. 
At equilibrium we have $P_+=1/2$, in agreement with standard statistical
mechanics considerations. 
Below $\epsilon_{dis}$, $P_+$ cannot  change in time because
the sign of $m_y$ remains the same for all trajectories.
Above $\epsilon_{dis}$, $P_+(t)$ can change in time.  Numerical results show
that $P_+(t)$ decays exponentially to the equilibrium value $1/2$ and that the 
time scale for reaching the equilibrium value is independent of the
initial probability distribution, $P_+(0)$, see  Fig.~\ref{ptcl}b) (dashed line).

A simple statistical model can explain the qualitative features of this
magnetic relaxation process.  
%The main assumptions of this model are
%that the system looses memory very fast and that magnetic reversal
%probability $\lambda$ is time independent. Such features could be a
%consequence of strong chaos: the system looses its memory due to
%sensitivity to initial conditions. 
Let us start with an ensemble of
$n$ initial conditions, of which $n_+$ with a positive magnetization
and $n_-$ with a negative magnetization, such that $n=n_++n_-$.
Assuming that $m_y$ can take only two values, $+$ and $-$, we can
write a pair of differential equation for the populations with
positive and negative magnetizations:
\begin{eqnarray}  
\label{nn}
\dot {n}_+& =& - \lambda  n_+ +\lambda  n_- \nonumber \\
\dot {n}_-& =& - \lambda  n_- +\lambda  n_+. \nonumber
\end{eqnarray}
Where $\lambda$ is the reversal probability per unit time defined above.
%%%%%%%%%%%%%%%%%%%%%%%%%%%%
Defining $P_+=n_+/n$, we can solve these equations, obtaining
\begin{equation}  
\label{pp}
P_+(t)-\frac{1}{2}= (P_+(0)-\frac{1}{2} )e^{-2 \lambda t}.
\end{equation}
$P_+(t)$ reaches the equilibrium value with a typical {\it relaxation
time} $\tau=1/(2 \lambda)$. 
% qui ho ancora sostituito twice con proportional
This simple model predicts a magnetic relaxation time, $\tau$, proportional to 
the average magnetization reversal time,
which checks pretty well with numerics 
(compare solid and dashed lines in Fig.~\ref{ptcl}b).
Therefore, hereafter we will use indifferently the two concepts.

\begin{figure}
\begin{center}
\includegraphics[scale=0.35]{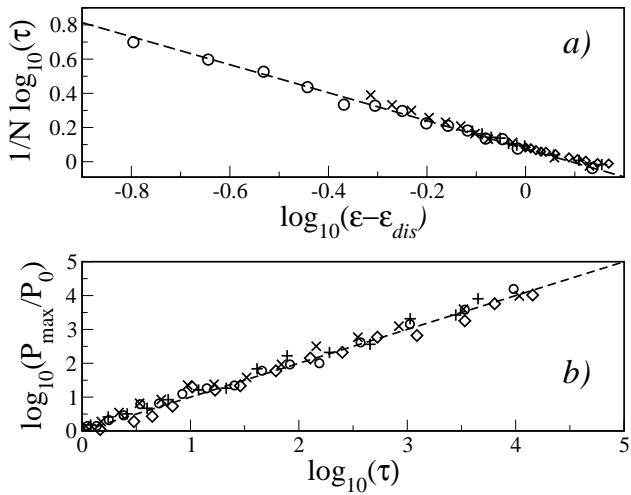}
\caption{ 
  Panel $a)$ shows the scaling of the magnetic relaxation time $\tau$
  for different values of $N$, for the case $B=0$, $J=3$.  The
  relaxation time has been computed starting from an ensemble of
  $100$ different initial conditions. The dashed line is the linear
  fit: $\frac{1}{N} log(\tau)=0.075-0.82
  log(\epsilon-\epsilon_{dis})$.  In panel $b)$ we show the
  times resulting from statistical mechanics, see Eq.~(\ref{b0pmax}),
  {\it vs} the 
  dynamically computed relaxation times, 
  for different $N$ values.  $P_{max}/P_0$, the main ingredient of formula
  (\ref{b0pmax}) is numerically determined for each energy density
  choosing $10^{6}$ points in a small energy interval. Symbols for
  the statistical times are as follows: $N=6$ (circle), $N=12$
  (cross), $N=24$ (plus), $N=48$ (diamond).  
  Dashed   line is $y=x$.
  }
\label{tau5}
\end{center}
\end{figure}

Analyzing the magnetic relaxation times for all energies in the range 
$(\epsilon_{dis},\epsilon_{stat})$, we find that they grow exponentially with
the number of spins for sufficiently large $N$, 
as expected for mean-field models.
More remarkable is the power law divergence of relaxation time at 
the non--connectivity threshold.
Numerical data are consistent with the following scaling law, 
\begin{equation}
\tau \sim \left( \frac{1}{\epsilon -\epsilon_{dis}} \right)^{\alpha N}~,
\label{tau}
\end{equation}
for which a theoretical justification will be given below.  Eq.~(\ref{tau})
is valid above the non-connectivity threshold and not too close to the
statistical  threshold $\epsilon_{stat}$. The comparison of this formula
with numerical results is shown in Fig.~\ref{tau5}.  

To explain and
substantiate these numerical findings, we now turn to an analytical
estimate of the relaxation times, based on statistical mechanics. In
Refs.~\cite{tau1,tau2}, on the basis of fluctuation theory
~\cite{landau,gri}, it has been argued that metastable states relax to
the most probable state on times proportional to $\exp (N\Delta s)$
where $N$ is the number of degrees of freedom and $\Delta s$ is the
specific entropic barrier. In our case $\exp (N\Delta s)$ is nothing
%modificato qui sotto
but $P_{max}/P_{0}$, where $P_{max}$ is the value
of $P(m_y)$ for the most probable value of $m_y$,
and $P_0=P(m_y=0)$. 
%%%%%%%%%%%%%%%%%%%%%%%%%%%%%%%%%%%%%%%%%%%%%%%%%%%%%
Thus, the exponential
divergence as a function of $N$ shown in Fig.~\ref{tau5}$a)$ is
consistent with Refs.~\cite{tau1,tau2}. These papers, however, did not
study the behavior of $\tau$ at fixed $N$ in the neighborhood of the
non--connectivity threshold. We perform this calculation in
Appendix~\ref{appb}, obtaining
\begin{equation}
\tau \sim 1/(\epsilon-\epsilon_{dis})^{\alpha N}~,
\label{tim33}
\end{equation}
with $\alpha=1$ generically, but $\alpha=3/4$ for $B=J=1$.  This result 
is qualitatively correct (power law divergence, exponent proportional to $N$), 
and quantitatively reasonable. 
%%% modificato qui:
Indeed, numerical simulations give $\alpha\simeq 0.82$
(instead of $\alpha=1$) for $J \gg B$ and $B=0$ (see Fig.~\ref{tau5}), 
and $\alpha\simeq 0.55$ (instead of $\alpha=3/4$) for $B=J=1$.  
%%%%%%%%%%%%%%%%%%%%%%%%%%%%%%%%%%%
We expect these qualitative features to be 
valid beyond the all-to-all coupling studied here, as it will be shown 
in Sec.~\ref{sec:general}.

The calculations to evaluate $P_{max}/P_{0}$ rely on
several approximations, the most doubtful being the large
$N$ assumption (as seen also in Section~\ref{sec:stat}). Hence, despite the
discrepancies in the exponents found above, the proportionality
between $\tau$ and $P_{max}/P_0$ may still be valid,
%% aggiunta questa frase
also for small $N$.  
%%%%%%%%%%%%%%%%
To test this
proportionality, we have calculated numerically the value of $P_{max}/P_0$, 
and we have found this value to be proportional to the 
relaxation time in any case.  
In particular, for the case $B=0$, we have found a very good fit setting
\begin{equation}
\tau  =  \frac{2}{J} \frac{P_{max}}{P_0}.
\label{b0pmax}
\end{equation}
The $P_{max}/P_0$ factor in this formula represents the probability to
cross the entropic barrier, and the $1/J$ factor can be heuristically
associated with the typical time scale of the system (for $B=0$ the
Hamiltonian is proportional to $J$). A deeper theoretical
justification of this formula should be obtained in view of its
success in describing the numerical results for different $N$ values
(see Fig.~\ref{tau5}$b)$). 

\begin{figure}
\begin{center}
\includegraphics[scale=0.35]{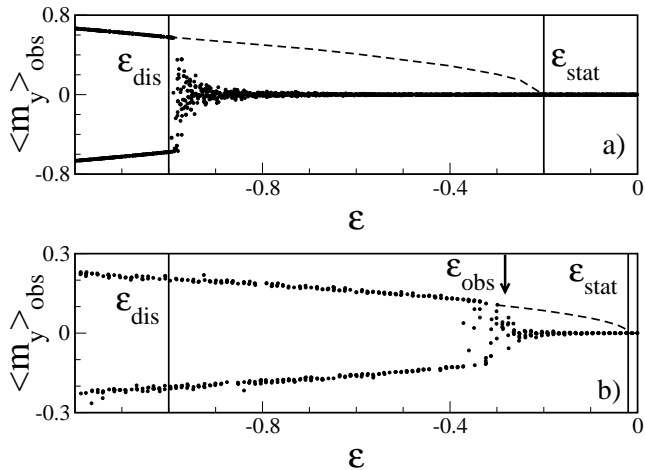}
\caption{
  Time average of $m_y$ over the observational time $\tau_{obs}$ {\it
    vs} $\epsilon$ for different number of particles (a) $N=5$, (b)
  $N=50$, with fixed $J=B=1$.  Each single point has been obtained
  taking the time average over the time intervals $\tau_{obs}=10^5$
  (a) and $\tau_{obs}=10^4$ (b).  Dashed  curves indicate the
  equilibrium value of $m_y$, obtained from statistical mechanics.
  Vertical lines represent the non-connectivity and the statistical
  threshold respectively.  The arrow in panel (b) indicates the energy value
  $\epsilon_{obs}$ of the {\it chaotic driven phase transition} due to the finite
  observational time.}
\label{fac}
\end{center}
\end{figure}

\subsubsection{Chaotic Driven Phase Transition}

Let us now answer the following question: if the measured
values of the magnetization are given by the time average of the
magnetization, for which energies the system will be found magnetized
and for which unmagnetized?\\

From Eq.~(\ref{tau}) 
and from the proportionality of the relaxation times and the reversal times, 
it is clear that the infinite time
average of the magnetization will be zero above the TNT
and different from zero below, due to the divergence of the
reversal time. Nevertheless, the conclusion is
  different for a finite observational time $\tau_{obs}$. In
  Fig.~\ref{fac} we show the time-averaged magnetization
$$
\langle m_y \rangle_{obs} = \frac{1}{\tau_{obs}} \int_0^{\tau_{obs}}\ dt
\ m_y (t)
$$
{\it vs} the specific energy $\epsilon$ for $N=5$ (Fig.~\ref{fac}a)
and $N=50$ (Fig.~\ref{fac}b) spins during a fixed observational time.
While in (a) $ \langle m_y \rangle_{obs}$ is zero just above
$\epsilon_{dis}$, in (b) it vanishes at a value $\epsilon_{obs}$
located between $\epsilon_{dis}$ and $\epsilon_{stat}$.  Indeed, if
$\tau_{obs} \gg \tau$, the magnetization has time to flip between the
two opposite states and, as a consequence, $\langle m_y \rangle_{obs}
\simeq 0$.  On the contrary, if $\tau_{obs} \ll \tau$ the
magnetization keeps its sign and cannot vanish during $\tau_{obs}$.
Defining an effective transition energy $\epsilon_{obs}$ from
$\tau_{obs} = \tau(\epsilon_{obs})$, one gets, inverting
Eq.~(\ref{tau}), the value indicated by the vertical arrow in
Fig.~\ref{fac}b. This is, {\it a posteriori}, a further
demonstration of the validity of Eq.~(\ref{tau}) for any $N$.

From a theoretical point of view, it is interesting to note that, for
any fixed $N$, if the fully chaotic regime persists down to
$\epsilon_{dis}$, $\epsilon_{obs} \to \epsilon_{dis}$ when $\tau_{obs}
\to \infty$.  On the other hand, in agreement with statistical
mechanics, for any finite $\tau_{obs}$, $\epsilon_{obs} \to
\epsilon_{stat}$ when $N\to \infty$.  This implies that the limits
$\tau_{obs} \rightarrow \infty$ and $N \rightarrow \infty$ do not
commute. From the above considerations it follows that if
$\tau_{obs}\to\infty$ at finite $N$, the threshold which distinguishes
between a magnetized energy region and an unmagnetized one is
$\epsilon_{dis}$ and not $\epsilon_{stat}$.  We can thus consider
$\epsilon_{dis}$ as the critical threshold at which a ``dynamical''
phase transition takes place: we call this transition
a {\it chaotic driven phase transition}.\\

Let us finally note that, usually, for long-range interactions, the
interaction strength is rescaled in order to keep energy 
extensive~\cite{kaz}. In our case this can be done setting $J=I/N$.  With
this choice of $J$ as $N \rightarrow \infty$, at fixed $I$, $J$
becomes much smaller than $B$, then a quasi--integrable regime sets in
and Eq.~(\ref{tau}) looses its validity (see Sec.~\ref{sec:quasint}).
The presence of the TNT is  therefore hidden.

\subsection{Quasi--integrable Regime}
\label{sec:quasint}

\begin{figure}
\begin{center}
\includegraphics[scale=0.35]{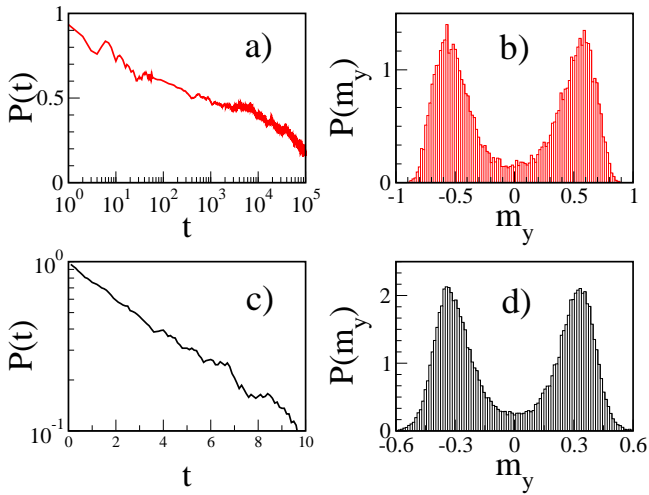}
\caption{(Color online)  Magnetic relaxation probability 
$P(t)=\frac{P_+(t)-1/2}{P_+(0)-1/2}$ and probability
distributions of $m_y$ in the {\it quasi--integrable} 
($J=0.5$, $\epsilon=-0.8$) 
(red panels a),b)), and {\it chaotic} ($J=3$,$\epsilon=-0.9$) 
(black panels c),d)) regimes. 
For both cases, $N=6$ and $B=1$.
The specific energies for the two cases have been chosen in such a 
way that $P_{max}/P_0$ is the same 
(see panels $b)$ and $d)$).
We compare $P(t)$ for the two different dynamical regimes.
In the fully chaotic regime, panel $c)$, $P(t)$ 
decays exponentially with time $t$, and the average relaxation time is of order $1$.
In the quasi--integrable regime, panel $a)$, the decay of $P(t)$ is  much slower (observe
the difference in the time axis scale).
}  
\label{pt2}
\end{center}
\end{figure}

In this Section we will give numerical evidence of the
quasi--integrable regime for $J < B$, in the energy region
between $\epsilon_{dis}$ and $\epsilon_{stat}$.  If the system dynamics
is not in a fully chaotic regime, there are important consequences
for reversal times. For instance reversal times 
%%ho aggiunto qt. frase.
strongly depend on initial conditions 
%%%%%%%%%%%%%%%%%%%%%%%%%%%%%%%%%%%%%%%%%%
and  Eq.~(\ref{tau}) looses its validity:

In Fig.~\ref{pt2} we consider a system 
with different interaction strengths $J$ in order to 
enter a quasi-integrable regime (a and b) and a chaotic one 
(c and d).
The energy in
the two cases has been chosen such that the entropic barrier is
roughly the same, see Fig.~\ref{pt2} $b)$ and $d)$. This means that,
from a statistical point of view, both systems are characterized by
roughly the same probability to jump over the barrier. Nevertheless,
as one can see in panels $a)$ and $c)$ of Fig.~\ref{pt2}, the behavior
of the probability $P(t)=\frac{P_+(t)-1/2}{P_+(0)-1/2}$
significantly differs in the two cases.  Such a big difference in the
statistical properties of magnetic reversal times can be explained
only by a drastic change in the dynamical properties of the system.
Indeed, while Fig.~\ref{pt2}a refers to a {\it quasi--integrable
  regime}, Fig.~\ref{pt2}c refers to a {\it fully chaotic regime}.
This cannot be explained by the different $J$ values, which, as we
have shown in the previous Section, have only a linear effect on the
reversal probability per unit time.

To better understand the origin of this quasi-integrable regime, it is
interesting to compare the dynamics obtained from the full
Hamiltonian (\ref{Hamiltonien2}) with the dynamics obtained from the
Mean-Field Hamiltonian (\ref{Hamiltonien3}).
Taking into account the conservation of the total angular momentum $M^2 =
m_x^2+m_y^2+m_z^2$, a change of variable maps (\ref{Hamiltonien3}) onto a 2
degrees of freedom Hamiltonian; the dynamics of the global
magnetization is then obviously integrable.

\begin{figure}
\includegraphics[scale=0.34]{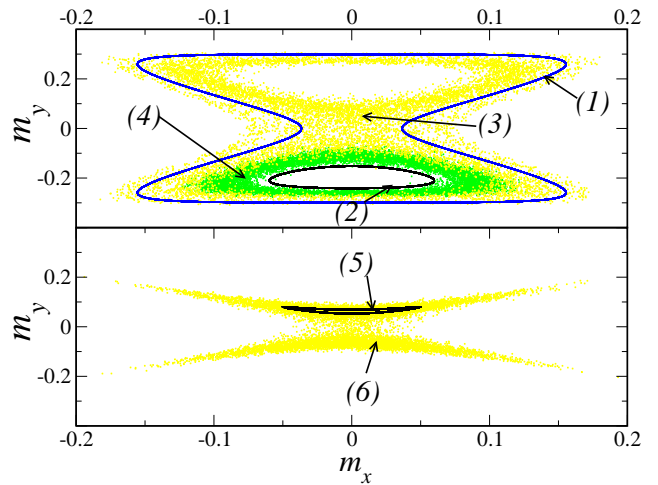}
\caption{(Color online)
Projections of the trajectories of $\vec{m}=(m_x,m_y,m_z)$ over the
$(m_x,m_y)$ plane for both the Mean-Field Hamiltonian (\ref{Hamiltonien3}) 
(heavy full lines : (1) blue and (2) black ) and the full Hamiltonian (\ref{Hamiltonien2})
(chaotically scattered dots (3) yellow and (4) green). 
In panel $a)$ ($J=0.2$, $\epsilon=-0.3$),
the system is in a quasi--integrable regime. The orbits of the full 
Hamiltonian remain close to the orbits of the Mean-Field Hamiltonian 
for all the integration time considered ($10^4$). Note also that in the some 
of the orbits cross the $m_y=0$ line, thus demagnetizing the system, and some do not.  
In panel $b)$ ($J=3$, $\epsilon=-0.32 $) the system is in a highly chaotic regime.  
In this case the orbit of the full Hamiltonian (6-yellow)  does not remain close to 
the one of the Mean-Field Hamiltonian (5-black) and covers most of the available phase--space.  
The integration time is, also in this case, $10^4$.}
\label{48jul}
\end{figure}

In Fig.~\ref{48jul} we show the projection of some trajectories on
the $(m_x,m_y)$ plane.  We considered the two different dynamical
regimes described above.  For definiteness, we vary $J$ but we choose
the specific energy in order to keep the same value of $P_{max}/P_0
\sim 20$.  Let us first discuss Fig.~\ref{48jul}a.  Dark lines
represent orbits of the Mean-Field Hamiltonian (\ref{Hamiltonien3}).  The
orbits of the macroscopic variable $\vec{m}=(m_x,m_y,m_z)$ cover tori,
since the Mean-Field Hamiltonian (\ref{Hamiltonien3}) is exactly integrable.
Nevertheless, trajectories display different features: while
trajectory $(1)$ crosses the line $m_y=0$, trajectory $(2)$ remains
confined in the negative ($m_y<0$) branch 
belonging to the same energy surface.
Two trajectories of
the full Hamiltonian (\ref{Hamiltonien2}) and the same initial
conditions are then considered, labeled by $(3)$ and $(4)$.  As one
can see these orbits stay for a long time sufficiently close to the
Mean-Field orbits.  Again, while $(4)$ displays a typical
``ferromagnetic'' behavior, $(3)$ is of ``paramagnetic'' nature.  Both
trajectories $(3)$ and $(4)$ have a positive maximal Lyapunov and are
therefore chaotic. Upon increasing $J$, and keeping the same value of
$P_{max}/P_0\sim 20$, we enter in the regime described by the lower panel in
Fig.~\ref{48jul}.  In this case, as above, we show the orbit $(5)$
of the Mean-Field Hamiltonian (actually a ``ferromagnetic'' one).  The
corresponding orbit of the full Hamiltonian, $(6)$, is still
characterized by a positive Lyapunov exponent and covers both
branches, $m_y>0$ and $m_y<0$, thus inducing the demagnetization of
the system.  What is important to stress is that in this case the
trajectories of the full Hamiltonian cover both the positive and the
negative magnetization branch on the same energy surface.  Having in
mind the mechanism that produces the transition to {\it global
stochasticity} in low-dimensional Hamiltonian
systems~\cite{Chirikov}, we can conjecture that invariant curves,
confining the motion, exist in the case of Fig.~\ref{48jul}a.  The
breakdown of these invariant curves signals the transition to a
globally chaotic motion.
Of course, characterizing such a breakdown is an
hard task, due to the high-dimensionality of the phase-space.

The determination of parameter regions in which the system is
quasi--integrable or fully chaotic is still an open question. We can
only make a few qualitative considerations. Let us consider
Hamiltonian (\ref{Hamiltonien2}); it contains the sum of two terms: a
mean-field integrable term plus the term $J/2 \sum_i
(S_i^y)^2-(S_i^x)^2$, which is responsible for the chaoticity of the
system.  The minimal specific energy of this term is $\epsilon_{chaos}
\sim -J/2$. We can thus suppose that for $\epsilon < \epsilon_{chaos}$
the quasi--integrable regime prevails, while for $\epsilon >
\epsilon_{chaos}$ a fully chaotic regime sets in.  Thus, in order to
have a fully chaotic regime in the energy region between
$\epsilon_{dis}$ and $\epsilon_{stat}$, it is necessary that
$\epsilon_{dis} > \epsilon_{chaos}$.  This is always the case if
$J>2B$ since for these values of $J$, $\epsilon_{dis} \sim
\epsilon_{chaos}$.  On the contrary, for $J<2B$ we expect a
quasi--integrable regime between $\epsilon_{dis}$ and
$\epsilon_{chaos}$, which should persist in the thermodynamic limit.

\section{Other Models}
\label{sec:general}
Till now, we have concentrated our analysis on a spin system with
all-to-all anisotropic coupling. The results obtained
concerning the TNT and the time scales
for magnetic reversal  can be
extended to more general situations. In this Section, we consider two
possible generalizations, and discuss how our results can be
extended to:  $i)$
distance dependent interactions and $ii)$ metastable states.

\begin{figure}
\includegraphics[scale=0.34]{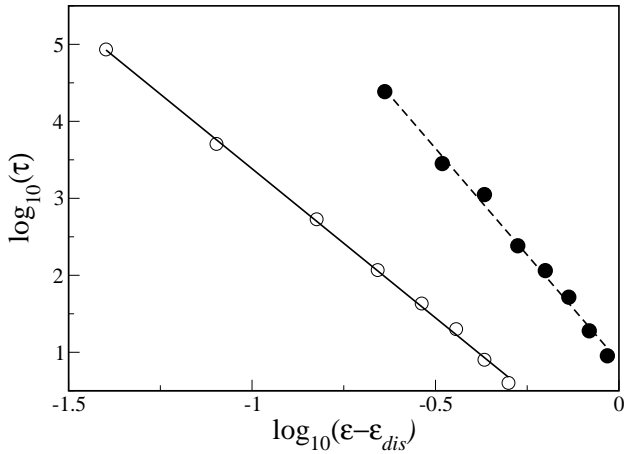}
\caption{Magnetization relaxation time $\tau$ {\it vs.} energy density 
for two different models.
(open symbols) a chain of $N=6$ Heisenberg spins with $R^{-\alpha}$ interaction and $\alpha=2$.
A best fit gives  $log(\tau) \sim -3.9 \ \log(\epsilon-\epsilon_{dis})$, with
$\epsilon_{dis} \simeq -0.71$. Here, $\epsilon_{min}=-1.083$.
(full symbols)  $3$-D cube with an additional spin at the center and nearest neighbor interaction. 
A best fit gives $log(\tau) \sim -5.5 \ \log(\epsilon-\epsilon_{dis})$, 
with $\epsilon_{dis} \simeq -1.33$. Here, $\epsilon_{min}=-20/9$.
}
\label{cube}
\end{figure}

$i)$ {\it Distance dependent coupling.} In Ref.\cite{brescia} it has been
considered a spin coupling which decays with the distance as
$R^{-\alpha}$.  It is possible to prove that in the $N\to\infty$
limit, for $\alpha<d$, a finite portion $r$ of the energy range
corresponds to a disconnected energy surface.  For $\alpha>d$, this
portion vanishes in the $N\to\infty$ limit. For finite $N$, however, a
well defined non connectivity threshold
$\epsilon_{dis}>\epsilon_{min}$ exists in both
the short and the long  case when the anisotropy
of the coupling induces an easy--axis of the magnetization.  Numerical
simulations support the conjecture that the behavior of the average
magnetization reversal  time is qualitatively similar to the $\alpha=0$
case. A power law divergence of the average reversal time when
$\epsilon$ approaches $\epsilon_{dis}$ is observed, see
Fig.~\ref{cube}.

More realistic models of micromagnetic systems include 3D clusters of
spins interacting only with their neighbors. Again, for large $N$, the
non connectivity threshold energy $\epsilon_{dis}$ converges to the
ground state energy $\epsilon_{min}$. However, for small clusters a
significant portion of the energy range corresponds to a disconnected
energy surface.  We have performed numerical simulations on a cluster
of 9 spins, arranged on a cube, with one spin in the middle. Each spin
of the cube interacts with its 3 neighbors and with the middle spin.
Fig.~\ref{cube} shows that the divergence of the magnetic relaxation
time close to $\epsilon_{dis}$ is again compatible with a power law.

\begin{figure}
\begin{center}
\includegraphics[scale=0.34]{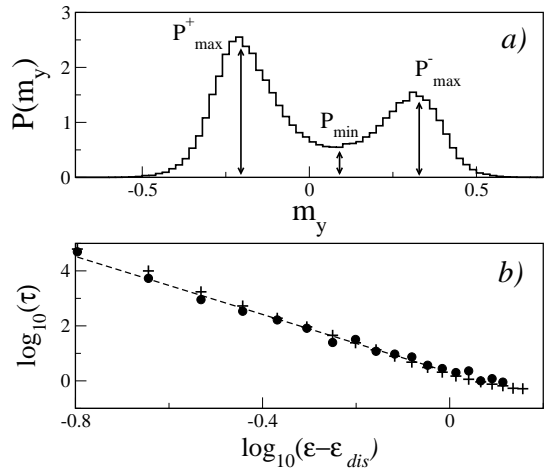}
\caption{Decay time from a metastable state. In panel $a)$ 
  $P(m_y)$ is shown for $\epsilon=-0.5$.  In panel $b)$ we show the
  power law divergence of the decay time from a metastable state (full
  circles).  To compute this decay time, $n=100$ initial conditions
  have been considered for each energy. The best linear fit
  $log(\tau)=0.309- 5.27 log(\epsilon-\epsilon_{dis})$ is also shown
  (dashed line).  The decay times computed from Eq.~(\ref{meta}) are
  shown as crosses. Apart from a deviation for high specific energy,
  the statistically computed decay times are in good agreement with
  the numerically computed ones (full circles). 
  Parameter values are: $N=6$, $J=3$,
  $B=0$, $B_y=1$}
\label{Meta1}
\end{center}
\end{figure}

$ii)$ {\it Metastable states.} The existence of the TNT has important
consequences for the decay time from metastable states. In order
to discuss this feature for a simple example, let us consider
Hamiltonian (\ref{ham}), adding a term, $B_y \sum S_i^y$,
 which contains a coupling to an external field
directed along the easy--axis of the magnetization.
In this case the non--connectivity threshold still exists (it has
the same value as before)  but the two peaks of $P(m_y,\epsilon)$ below
$\epsilon_{stat}$ do not have the same height, see Fig.~\ref{Meta1}a.
Thus, we can consider the time needed to reach the equilibrium value
of the magnetization if we start from a metastable state.
%% modificato qui
Below
$\epsilon_{dis}$ metastable states becomes stable for any finite $N$. 
%%%%%%%%%%%%%%%%%%%%%%%%%%%%%%%%%%%%%%%%%%%%%%%%%%%%%%%%%%%%%%%%%%%%%%
Above $\epsilon_{dis}$ the decay time diverges at
$\epsilon_{dis}$ as a power law, see Fig.~\ref{Meta1}b.
This decay time can be estimated from the statistical properties of the system.
Indeed, employing the same simple model described in Sec.~(\ref{subtime}),
we can evaluate the decay time scale.
Denoting by $P^+_{max}$, $P^-_{max}$ and $ P_{min}$ the probabilities
of the thermodynamic stable, metastable and unstable state, respectively (see 
Fig.~\ref{Meta1}a), and setting  $\lambda_\pm=P^\pm_{max}/P_{min}$,  
and $q=P^+_{max}/P^-_{max}$ we get 
the following  estimate of the decay time: 
\begin{equation}
\tau \sim \frac{q}{1+q} \frac{P^-_{max}}{P_{min}}. 
\label{meta}
\end{equation}
The good agreement of this estimate with the computed decay times is 
shown by the crosses in Fig.~\ref{Meta1}b.

%%%%%%%%%%%%%%%%%%%%%%%%%%%%%%%%%%%%%%%%%%%%%%%%%%%%%%%%%%%%%%%%%%%%%%%%
\section{Conclusions}

Anisotropic Heisenberg spin models with all-to-all coupling show a
Topological Non-connectivity Threshold (TNT) energy~\cite{jsp}.  Below
this threshold the energy surface splits in two
components, with opposite easy-axis magnetizations,
and ergodicity is broken, even with a finite number $N$ of spins. In
the same model, a second order phase transition is present, at an
energy higher than the TNT energy. We have fully characterized this
phase transition in the microcanonical ensemble, using a newly
developed method, based on large deviation theory~\cite{thierry}.  For
energies in the range between the TNT and the phase transition,
magnetization randomly flips if certain strong chaotic motion features
are present: the statistics of magnetization reversals is Poissonian.
Based on the knowledge of the microcanonical entropy
as a function of both energy and magnetization, we have derived a
formula for the average magnetic reversal time, which is valid in the
large $N$ limit. This formula agrees well with numerical results.
The formula also
predicts a power law divergence of the mean reversal time at the TNT
energy, which is also well verified in  numerical
experiments.  The exponent of the power-law divergence is also in
reasonable agreement,  although finite $N$ effects are 
quantitatively  important.

Finally, we have shown that all these features (presence of a TNT,
power-law divergence of the reversal time, etc.) are not limited to
systems with all-to-all coupling. The phenomenology is qualitatively
the same for anisotropic Heisenberg spin models with distance
dependent interactions and  for small clusters
of Heisenberg spins with nearest neighbor coupling.
We also considered systems where metastable states are present.
In this case, while below the TNT they are trapped, above it their
decay time diverges as a power law at the TNT.
Therefore, we conjecture that the power--law divergence of
the magnetic reversal time may be a universal signature of the
presence of the TNT, which is a generic feature of systems with
long--range interactions or small systems for which the range of the
interaction is of the order of system size, if the anisotropy of the coupling 
is such to determine an easy-axis of the magnetization.

\section{Acknowledgments}

SR acknowledges financial support under the contract COFIN03 {\it
  Order and Chaos in Nonlinear Extended Systems}. Work at Los Alamos
National Laboratory is funded by the US Department of Energy. We
thank T.~Dauxois, E.~Locatelli, F.~Levyraz, F.~M.~Izrailev and
R.~Trasarti-Battistoni for useful discussions.

\appendix

\section{Minimum energy}
\label{appa}
In this section we find the minimum of the Mean-Field  Hamiltonian 
(\ref{Hamiltonien3}):
\begin{equation}
\epsilon= m_z +\frac{I}{2}(m_x^2 - m_y^2).
\label{sie}
\end{equation}
It is sufficient to find the absolute minimum of
$$m_z-(I/2)m_y^2,$$ and verify that it satisfies $m_x=0$.
Taking derivatives
\begin{equation}
\begin{array}{lll}
\displaystyle
\frac{\partial N\epsilon}{\partial \phi_i} &= I m_y \sin \theta_i \cos \phi_i = 0\\
&\\
\displaystyle
\frac{\partial N\epsilon}{\partial \theta_i} &= -\sin \theta_i -I m_y 
\cos \theta_i \sin \phi_i = 0~.\\
\label{conn1}
\end{array}
\end{equation}
one gets two kinds of solutions (both with $m_x=0$):
\begin{itemize}
\item [1)] $\theta_i=\pi$ and $\phi_i = 0,\pi$
\item [2)] $\phi_i = \pm \pi/2$ and $\tan \theta_i = \pm Im_y.$
\end{itemize}
Let us define $N n_1 $ the number of solutions of type 1) and $N n_2$
the number of solutions of type 2) so that $n_1+n_2=1$.
Since $m_z = -n_1 - n_2 \cos\bar{\theta}$ and $m_y = \pm n_2 \sin\bar{\theta}$
where $\bar{\theta}$ is the solution of type 2), condition 2) is equivalent to
$\cos\bar{\theta}=1/In_2$.
Therefore, when $In_2< 1$ the set defined from 2) is empty and only solutions
in the class 1) can be obtained.
It is also easy to find the expression for the energy in terms
of $ 1/I \leq n_2 \leq 1 $:
\begin{equation}
\epsilon = -1-\frac{1}{2I}+n_2 -\frac{I}{2}n_2^2.
\label{mesm}
\end{equation}
Minima must be sought among the extrema so that when $n_2=1$ then
$e_{min}=-1/2I -I/2$ and when $n_2=1/I$ then $\epsilon_{min}=-1$. In
terms of $I$, one then has
\begin{equation}
\epsilon_{min}=\left\{
\begin{array}{lll}
&-1/2I-I/2 \ {\rm~~~for} \qquad I\geq 1\\
&\\
&-1 \quad{\rm~~~~~~~~for} \qquad I < 1.
\end{array}
\right.
\label{miesm}
\end{equation}
From Eq.~(\ref{miesm}) we have (\ref{cri}),  using transformations 
in (\ref{trasf}).

\section{Critical exponents}
\label{appb}

In this section, we study the divergence of the reversal  time 
for $\epsilon \to \epsilon_{dis}^+$, at fixed $N$. Let us 
assume that it is given by
$$
\tau \simeq \max_{m_y} P(m_y,\epsilon)/P(m_y=0,\epsilon)~,
$$ 
we show that
\begin{equation}
\tau \simeq \frac{1}{(\epsilon-\epsilon_{dis})^{\alpha N}}~,
\label{scal}
\end{equation}
with $\alpha$ a constant independent of $N$; we find $\alpha=1$ or
$\alpha=3/4$, depending on the parameters of the Hamiltonian.\\

First, we note that although $\max_{m_y} P(m_y,\epsilon)$ increases
exponentially with $N$ at fixed $\epsilon$, it does not change much at
fixed $N$ when $\epsilon \to \epsilon_{dis}^+$; the behavior of $\tau$
is dominated by the value of $P(m_y=0,\epsilon)\propto
e^{-Ns(m_y=0,\epsilon)}$. The problem is then reduced to the
computation of $s(m_y=0,\epsilon)$.

Before turning to the actual calculation of $s(m_y=0,\epsilon\to
\epsilon_{dis})$, we consider the following problem, which will be
useful later. Let us consider the random variable
$x$ in $[-1,1]$ with distribution
$\rho(x)$; we call $m=\sum_{i=1}^N x_i/N$, and ask the question: what
is the behavior of $P(m=1-\delta),~\delta<<1$, for fixed $N$
reasonably large? Using Cram\'er's theorem, we write
\begin{equation}
\label{eq:psiapp}
\Psi(\lambda) = \int_{-1}^{1} \rho(x) \exp(\lambda x) dx
\end{equation}
and
\begin{equation}
\label{eq:entapp}
s(m)=-\sup_\lambda \left( \lambda m -\ln \Psi(\lambda) \right)~.
\end{equation}
$m\to 1$ implies $\lambda \to \infty$, so the behavior of $\rho$ close
to $x=1$ dominates (\ref{eq:psiapp}). We write $\rho(x)\sim
a(1-x)^\gamma$, close to $x=1$, with $\gamma>-1$. Then for
$\lambda\to\infty$,
$$
\Psi(\lambda) \sim  a\frac{e^\lambda}{\lambda^{\gamma+1}}
\int_0^\infty u^\gamma e^{-u} du~.
$$
Then the maximizing $\lambda$ in (\ref{eq:entapp}) is given by
$m=1-(\gamma+1)/\lambda$. Substituting into (\ref{eq:entapp}), we get
$s(m=1-\delta)\sim (\gamma+1)\ln\delta$, and finally,
$P(m=1-\delta)\sim \delta^{N(\gamma+1)}$.\\

We now apply this result to the easiest case, the simplified
Hamiltonian $H=N(m_z+I(m_x^2-m_y^2)/2)$. The threshold is
$\epsilon_{dis}=-1$ (we consider the case $I>1$); we write
$\epsilon=-1+\delta$, with $\delta<<1$. We want to compute the entropy
$s(m_x,m_y=0,m_z=-1+\delta-Im_x^2)$. Noticing that a small $\delta$
implies a small $m_x$, we simplify the calculation to
$s(m_x=0,m_y=0,m_z=-1+\delta)$. We now use once again Cram\'er's
theorem.
\begin{eqnarray}
\Psi(\lambda_x,\lambda_y,\lambda_z)&=&\frac{1}{4\pi}\int_0^\pi~\sin
\theta d\theta \int_0^{2\pi} d\phi \exp(\lambda_x \sin \theta \cos
\phi) \nonumber \\
&& \exp(\lambda_y \sin \theta \sin \phi+\lambda_z \cos \theta)~. 
\end{eqnarray}
Then $s$ is given by:
\begin{eqnarray}
s(m_x,m_y,m_z) &=& -\sup_{\lambda_x,\lambda_y,\lambda_z}
\left[m_x\lambda_x+m_y\lambda_y+m_z\lambda_z \right. \nonumber \\
&& \left. -\ln \Psi(\lambda_x,\lambda_y,\lambda_z)\right]~.
\end{eqnarray}
For $m_x=m_y=0$, the maximizing $\lambda_x$ and $\lambda_y$ are
found to vanish. Thus, the problem reduces to calculating
$s(m_z=1-\delta)$ (using also the symmetry $m_z \to -m_z$). Recalling
that $m_z=\langle \cos \theta \rangle$, with $\theta$ the latitude of
a point taken randomly on the sphere with uniform probability. This is
equivalent to saying that $m_z=\langle q \rangle$, with $q$ a random
variable uniformly distributed between $-1$ and $1$. Using the general
result derived above with $\gamma=0$, we find that $\tau \sim
(\epsilon-\epsilon_{dis})^{\alpha N}$, with $\alpha=1$.\\

We turn now to the complete Hamiltonian, with $B=0$,
$H=N(I(m_x^2-m_y^2)/2-J\Delta/2)$. The threshold is now
$\epsilon_{dis}=-J/2$. We set $\epsilon=-J/2~(1-\delta)$. Noticing
that again, a small $\delta$ implies a small $m_x$, we compute
$s(m_x=0,m_y=0,\Delta=1-\delta)$, in the limit of small $\delta$.
$\Delta$ is defined as $\langle \sin^2\theta\cos 2\phi \rangle$, for
$\theta$ and $\phi$ coordinates of points taken randomly on the sphere
with uniform probability. Again, this is equivalent to saying that
$\Delta=\langle q \rangle$, with $q$ now having a non uniform
distribution $\rho(q)$ in $[-1,1]$. However, $\rho(q)$ tends to a
constant value as $q\to 1^-$ (the calculation is detailed at the end
of the appendix), which means $\gamma=0$; thus Eq.(\ref{scal}) holds,
again with $\alpha=1$. The conclusion is the same for all $B\neq J$.\\

Finally, we consider now the case $B=J=1$. Then
$H=N(m_z+N(m_x^2-m_y^2)/2-\Delta/2)$. Setting $\epsilon=-(1-\delta)$,
we want to compute $s(m_x=0,m_y=0,m_z-\Delta/2=1-\delta)$. Calling
$M=m_z-\Delta/2$, we have $M=\langle q \rangle$, with $q=\cos \theta
-(\sin^2\theta \cos 2\phi)/2 $ a random variable in $[-1,1]$ with
distribution $\rho(q)$. The calculations in the next paragraph show
that $\rho(q)$ diverges at the boundary like $(1-q)^{\gamma}$, with
$\gamma=-1/4$; thus Eq.(\ref{scal}) still holds, now with
$\alpha=3/4$.\\

\emph{Derivation of the exponent $\gamma$}~:\\

1. $\mathbf{B=0}$: we need to compute the distribution
$P(y=\sin^2\theta \cos 2\phi)$, close to $y=1$. We have, with $u=\cos
\theta$:
\begin{equation}
P(y)= \int_{-1}^{1} du \int_0^{2\pi} d\phi~\delta_D \left( y-(1-u^2)\cos
  2\phi \right)~,
\end{equation}
where $\delta_D(x)$ is the Dirac delta function.
Writing $y=1-\delta$, with $\delta\ll 1$, we see that only the values
of $u$ such that $u^2<\delta$ contribute. 
Integrating  over $\phi$ we obtain:
\begin{equation}
P(1-\delta)= 4 \int_{-\sqrt{\delta}}^{\sqrt{\delta}}  
\frac{du}{2((1-u^2)^2-(1-\delta)^2)^{1/2}}~;
\end{equation}
the factor of $4$ in front comes from the four values of $\phi$ that
contribute the same amount to $P$. Expanding the denominator,
neglecting order $\delta^2$ terms and performing the change of
variable $u=\sqrt{\delta}t$, we get:
\begin{equation}
P(1-\delta)\simeq \int_{-1}^{1} \frac{\sqrt{2} dt}{\sqrt{1-t^2}}~. 
\end{equation}
Since this last integral does not depend on $\delta$ and has a
finite value, we conclude that $\gamma=0$ for $B=0$.\\

2. $\mathbf{B=J=1}$: we need to compute now $P(y=u-(1/2)(1-u^2)\cos
   2\phi)$ close to $y=1$. This reads:
\begin{equation}
P(y)= \int_{-1}^{1} du \int_0^{2\pi} d\phi \delta_D \left(
  y-u+\frac{1}{2}(1-u^2)\cos 2\phi \right)~.
\end{equation}
Solving for $\phi$ inside the delta function, we get
\begin{equation}
\cos 2\phi =\frac{2(u-y)}{1-u^2}~.
\end{equation}
This time, only the values of $u$ close to $u=1$
contribute to $P$; thus, we write $u=1-s$, $y=1-\delta$. From the
inequalities $-1\leq \cos 2\phi \leq 1$, we get, neglecting terms of
order $s^2$, $\delta/2 \leq s \leq \sqrt{2\delta}$. Integrating the
delta function over $\phi$, we have the expression for $P$:
\begin{equation}
P(1-\delta) \simeq 4 \int_{\delta /2}^{\sqrt{2\delta}}  
\frac{ds}{\left( (2s-s^2)^2-4(\delta-s)^2 \right)^{1/2}}~;
\end{equation}
the change of variable $t=2s/\delta$ yields
\begin{equation}
P(1-\delta)\simeq  \int_{1}^{\sqrt{8/\delta}}  
\frac{dt}{\sqrt{t-1}}~.
\end{equation}
This integral converges close to $t=1$; it diverges however at large
$t$, like $t^{1/2}$; since $t$ diverges as $\delta^{-1/2}$, we finally
get $\gamma=-1/4$.\\

3. {\bf General case} $\mathbf{B \neq J}$: we do not detail here the
calculations, which are similar to those above. As soon as $B \neq J$,
the result is $\gamma=0$, and thus $\alpha=1$.\\

%%%%%%%%%% fin de l'appendice modifie... %%%%%%%%%


\begin{thebibliography}{}

%\bibitem{Nat1} P. Gambardella et al., Nature Vol.416, 301 (2002).

\bibitem{Gross}D. H. E. Gross \emph{Microcanonical
      Thermodynamics: Phase Transitions in Small Systems}, Lecture
      Notes in Physics {\bf 66}, World Scientific, Singapore, 2001.

\bibitem{chaos} F. Borgonovi, G. Celardo, F. M. Izrailev, and G. Casati
Phys. Rev. Lett., {\bf 88}, 054101 (2002); V. V. Flambaum and F. M. Izrailev,
Phys. Rev. E, {\bf 56}, 5144, (1997); F. Borgonovi and F. M. Izrailev,
Phys. Rev. E, {\bf 62}, 6475 (2000);  F.~Borgonovi, I.~Guarneri,
F.~M.~Izrailev and G.~Casati, Phys. Lett. A, {\bf 247}, 140 (1998).

\bibitem{nano} M.~Hartmann, G.~Mahler and O.~Hess. Phys. Rev. Lett.
  {\bf 93}, 80402 (2004).

\bibitem{jsp} F.~Borgonovi, G.~L.~Celardo, M.~Maianti, E.~Pedersoli,
J. Stat. Phys., {\bf 116},  1435 (2004).

\bibitem{brescia} F.~Borgonovi, G.~L.~Celardo, A.~Musesti, 
R.~Trasarti-Battistoni and P.~Vachal, cond-mat/0505209.

\bibitem{cornell} L.~Q.~English, M.~Sato and A.~J.~Sievers, Phys. Rev. B , {\bf 67}, 24403 (2003);
M.~Sato et al., Jour. of Appl. Phys.,  {\bf 91} , 8676 (2002).

\bibitem{thebibble} T.~Dauxois, S.~Ruffo, E.~Arimondo, M.~Wilkens Eds.,
Lect. Notes in Phys.,  {\bf 602},  Springer (2002).

\bibitem{thierry} R.~S.~Ellis, Physica D, {\bf 133}, 106 (1999);
  J.~Barr\'e, F.~Bouchet, T.~Dauxois, S.~Ruffo, \emph{J. Stat. Phys.}
  {\bf 119}, 677 (2005); J.~Barr\'e, Phd Thesis, ENS-Lyon, (2003).

\bibitem{num} We simulated the dynamic of the system through a 
Runge Kutta $4$-th order integrator.

\bibitem{phd} G.~L.~Celardo, PhD Thesis,  Univ. of Milan (2004).

\bibitem{odd}  For $J>B$ the results for odd and even $N$ coincide 
up to ($1/N$) corrections.

\bibitem{chud} E.~M.~Chudnovsky and J.~Tejada, {\it Macroscopic Quantum
Tunneling of the Magnetic Moment}, Cambridge University Press,
(1998).

\bibitem{Dembo} A.~Dembo, O.~Zeitouni, \emph{Large Deviations Techniques and
Applications}, (Springer, Berlin, 1998).

\bibitem{pmy} Note that the same distribution can be obtained, 
in a fully chaotic regime, from the time series of $m_y(t)$.

\bibitem{tel} Sh. Kogan, {\it Electron Noise and Fluctuations in Solids}
Cambridge Univ. Press., Cambridge  (1996).

\bibitem{tau1} M.~Antoni, S.~Ruffo and A.~Torcini, Europhys. Lett.,
{\bf 66}, 645 (2004).

\bibitem{tau2} P.~H.~Chavanis and M.~Rieutord Astronomy and Astrophysics, 
{\bf 412}, 1 (2003); P.~H.~Chavanis, astro-ph/0404251.

\bibitem{landau} L.~D.~Landau and E.~M.~Lifshitz, {\it Statistical Physics} 
Pergamon Press, Oxford (1985).

\bibitem{gri} R.~B.~Griffiths, C.~Y.~Weng, and J.~S.~Langer, Phys. Rev.,
{\bf 149}, 1 (1966). 

\bibitem{kaz} M.~Kac, G.~E.~Uhlenbeck and P.~C.~Hemmer, J. Math.
  Phys., {\bf 4}, 216 (1963).

\bibitem{Chirikov} B.V.Chirikov Phys. Rep., {\bf 52}, 253 (1979).

%\bibitem{prep3} F.Borgonovi, G.L.Celardo, A.Musesti, R.Trasarti-Battistoni,
%                cond-mat/0505209.



%\bibitem{Palmer} R.~G.~Palmer, Adv. in Phys., {\bf 31}, 669 (1982).
  
%\bibitem{tau3} P.~Hanggi, P.~Talkner, M.~Borkovec, Rev. Mod. Phys.
%  {\bf 62}, 2 (1990).


%\bibitem{Arnold} V.I. Arnol'd, DAN. USSR, {\bf 156}, 9 (1964),
%A.J. Lichtenberg, M.A. Lieberman, {\it Regular and Stochastic Motion},
%Springer-Veriag (1983).



\end{thebibliography}
\end{document}